\documentstyle[12pt]{article}
\def\d{{\rm d}}
\begin{document}
\begin{flushright}
UT-711

OCHA-PP-49

August, 1995
\end{flushright}
\vspace{1.0cm}
\begin{center}
{\LARGE \bf Corrections to Angular Ordering in Multiple Hadroproduction}
\end{center}
\vspace{0.5cm}
\begin{center}
{\large K. $\!$Kimura}\\
Department of Physics, Ochanomizu University\\
Otsuka, Bunkyo-ku, Tokyo 112, Japan\\
\vspace{0.2cm}
{\large M. $\!$Kitazawa}\\
Department of Physics University of Tokyo\\
Hongo, Bunkyo-ku, Tokyo 113, Japan\\
\vspace{0.2cm}
{\large K. $\!$Tesima}\\
Department of Physics, Ochanomizu University\\
Otsuka, Bunkyo-ku, Tokyo 112, Japan
\end{center}
\date{}
\vspace{2.0cm}

\begin{abstract}
One of the key ideas in describing the multiparticle production
at high energies in the perturbative QCD is
the angular ordering in successive soft-gluon emission.
We analyse the angular distribution of particles in a jet,
and investigate the corrections to the angular ordering.
At a small angle from the jet direction,
the angular ordering is exact at the next-to-leading order.
At a large angle, the angular ordering overestimates the angular
particle density by a factor $(1-\gamma\ln\,\cos^2(\theta/2))$,
where $\gamma$ is the anomalous dimension of the multiplicity.
The $O(\gamma)$ correction
restores the boost invariance of the cross section.
\end{abstract}

\newpage

\section {Introduction}

The modified leading-log approximation
(MLLA)\cite{bas}\cite{dok2}\cite{dok1} is the formalism
to systematically resum the double- and single-logarithms
occurring in analysing the multiparticle production
at high energies in the perturbative QCD.
After the resummation, the perturbation theory is reorganised
in power series of $\gamma$, where $\gamma$
is the anomalous dimension of
the multiplicity: $\gamma\simeq\sqrt{3\alpha_s/(2\pi)}$.
The hypothesis of the local parton-hadron duality
(LPHD)\cite{ama}\cite{dok2}\cite{azi},
on the other hand, predicts that the partonic
particle flow at high energies
(calculated in the perturbation theory)
is proportional to the hadronic flow
(to be observed in the experiments).
This framework (MLLA+LPHD) had remarkable successes
particularly in predicting the shape of
the particle spectrum in the e$^+$e$^-$-annihilation,
in excellent agreement with the experimental
data\cite{dok2}\cite{azi2}\cite{opa1}.

One of the key ideas underlying the MLLA is
the angular ordering (AO)\cite{mue1}
in the successive soft-gluon emission.
Because of the presence of the soft singularity in QCD,
the majority of the particles originate from the soft\footnote{
By \lq\lq soft" we mean at much lower energies than the jet energy,
but still at much higher than the inverse of the hadron size
so that the effective QCD coupling for their emission is
small} gluons:
An energetic parton emits soft gluons,
the soft gluons in turn emit softer gluons, and so on.
The cascade of soft gluons thus gives rise to large multiplicities
at high energies.
A large-angle soft gluon, however, cannot resolve the transverse
polarisation in a jet at scales smaller than the inverse of the
transverse momentum of the soft gluon.
Its emission, therefore, does not take place
independently from each individual parton in a narrow jet.
A large-angle soft gluon is emitted from a narrow jet
as though from a single energetic particle
with the net charge of the jet
(\lq\lq coherence")\cite{azi3}\cite{dok4}.
It implies that only those gluons that are emitted
{\it at small angles}
participate the cascade of partons
and contribute to the major part of the multiplicity (AO).
The coherence is also the origin of the characteristic bell-shape
of the particle spectrum observed
in the e$^+$e$^-$-annihilation\cite{opa1}.

One of the important applications of the MLLA
is the coherent shower Monte Carlo simulation
of the multiparticle production\cite{mar}\cite{gus}.
{}From the theoretical points of view,
the simulation is meant to give the total particle
multiplicity and the particle
spectrum correctly (at the next-to-leading order of MLLA).
The coherent shower Monte Carlo programs seem to be able to
reproduce universally well the events in the e$^+$e$^-$-annihilation.

While the AO is a powerful picture which enables us to organise the
parton cascade as a simple Markov process,
it is as well important to closely examine its limitation.
The picture of coherent soft-gluon emission may appear to be a
rough approximation which holds only when a soft gluon
is emitted from a narrow jet.
When the jet is not necessarily very collimated,
a more quantitative analysis is required.
It has been known that the AO is correct
at the next-to-leading order ($O(\alpha_s)$)
for angle-averaged quantities (exact angular ordering).
For such quantities, therefore, the correction
due to the transverse polarisation of a jet
(\lq\lq dipole correction")
appears at the next-to-next order ($O(\gamma^2)$).
This type of correction was evaluated at the lowest order
(two-gluon emission cross section
from $q\bar q$, $O(\alpha_s^2)$) by Dokshitzer
et al.\cite{dok2}\cite{dok1} in the planar gauge.
They made use of the result to evaluate the correction
to the total multiplicity.

In the quantities which depend on the angles, on the other hand,
the dipole correction can appear
at the next-to-leading order c.
In this article, we shall closely examine the angular distribution
of particles in a jet.
At the leading order (the double-log approximation),
the direction of the registered particle may be identified
with the direction of the subjet of the primary gluon
(the gluon directly emitted from the high-energy parton)
to which it belongs.
Owing to the factorisation of the soft-gluon emission amplitude,
the angular density of particles is given at this order
as the product of two amplitudes\cite{smi}:
the probability of the emission of a primary gluon at the angle
of the registered particle,
and the particle multiplicity in the subjet with an opening angle
of the size of the emission angle of the primary gluon.
This approximation is justified under the strong AO.

At higher orders,
we should take into account the difference in direction
between the primary gluon and the registered particle.
The density of the secondary gluon, though sharply peaked
in the direction of the primary gluon,
spreads inside the angular cone of the size of the emission angle
of the primary gluon.
One might therefore expect that the sharp leading-order peak
of the distribution in the direction of a jet may be
somewhat \lq\lq smeared" by the spread of the secondary gluon.
We shall find, however,
that such smearing of the peak does not take place.
Namely, the AO is exact at small angles.
As for the particle density far from the jet direction,
on the other hand,
the AO has to be modified at the next-to-leading order.
We shall find, however, that the correction is due to the
lack of the Lorentz covariance in the angular variables.
If we formulate the MLLA in terms of Lorentz invariants,
as in \cite{tes2}\cite{mun},
we do not have the dipole correction.

In the next section, we review the AO formalism.
Though the formulae presented in this section are all well-known,
we include them in order to make the text self-contained.
We first derive the AO for the angle-averaged cross section for
one-gluon emission $(O(\alpha_s))$.
The result is generalised to all orders in $/alpha_s$.
The AO is exact at the next-to-leading order ($O(\alpha_s)$).
The AO, however, is by no means obvious
for the quantities which depend explicitly on the angle.

In sect.3, we discuss how much the exact soft-gluon emission amplitudes
are modified by the approximation of the AO.
We analyse in detail the Feynman diagrams for two-gluon emission
(at the tree level), and examine the possible
next-to-leading correction
in the one-particle-inclusive (1PI) cross section.
The result is generalised in sect.4 beyond
the fixed order perturbation:
i.e. $\!$the large logarithms are resummed
to all orders in $\alpha_s$ (MLLA).

We find that there is no correction to the AO at a small angle.
This is owing to the symmetry of the cross section under the exchange
between the primary and the secondary gluons:
After integrating the two-gluon emission cross section over the
momentum of the primary gluon, we are left with the same
angular form for the secondary gluon as that for the primary
gluon\footnote{This is the case only for a massless quark.
When the quark jet is from a heavy quark,
there is a substantial correction to the AO at the forward direction.
See ref\cite{kit1}.}.

In sect.5, we evaluate the correction at a large angle,
where the dipole correction appears at the next-to-leading order.
We find that the correction reduces the angular density
by a factor $(1+\gamma\ln\,\cos^2(\theta/2))$.
The correction restores the invariance of the cross section
under the boost in the jet direction.

\section{The Angular Ordering}

In this section,  we review the AO
in the successively softer gluon
emission\cite{mue1}\cite{azi}\cite{azi2}\cite{dok4}.
First, we consider one soft-gluon emission from the $q\bar{q}$-pair.
A simple form of the AO, which is exact at the next-to-leading order,
is obtained after dividing the factorised
soft-gluon emission amplitude into two terms and performing azimuthal
integration in each term.
Then we analyse the soft-gluon emission from a gluon.
We generalise the result to the successive soft-gluon emission
at an arbitrary order in $\alpha_s$.

\subsection{One-Gluon Emission}

The lowest order amplitude for one soft-gluon emission
from $q\bar{q}$-pair (Fig.1) is essentially the same
as the well-known soft-photon emission amplitude in QED,
with the QCD coupling $\alpha_s$
in place of the fine structure constant $\alpha$.
In the soft limit, the gluon emission
amplitude factors out from the lowest order $q\bar q$-pair production
cross section $\sigma_0$(e$^+$e$^-\rightarrow q\bar q$):
\begin{eqnarray}
\d\sigma(e^+e^-\to q\bar{q}g)
&=& \frac{4\pi\alpha_s C_F\, \d^3 k}{(2\pi)^3 2k}
\frac{2p_1 \cdot p_2}{(k \cdot p_1)(k \cdot p_2)}\, \sigma_0
\nonumber\\
\nonumber\\
&=& \frac{C_F \alpha_s\,\d k \,\d \Omega}{2\pi^2 k}
\frac{1- \cos \theta_{p_1p_2}}{(1- \cos \theta_{kp_1})
(1- \cos \theta_{kp_2})}
\, \sigma_0
\end{eqnarray}
where $p_1$, $p_2$ and $k$ are the momenta of the quark,
the antiquark and the gluon respectively.
$\theta_{kp_1}$ ($\theta_{kp_2}$) is the angle
between the directions of
motion of the gluon and the quark (antiquark),
$\theta_{p_1p_2}$ is the angle between the quark and antiquark,
and $\Omega$ is the solid angle of $\vec k$.
The quark colour charge $C_F$ is defined by
$\displaystyle\sum_a(\tau_a \tau_a)_{\alpha \beta}=
C_F\delta_{\alpha \beta}$,
where $\tau_a$ is the quark colour matrix.
$C_F= (N_c^2-1)/2N_c$ for SU($N_c$) ($C_F=4/3$ with $N_c=3$).

The angular factor in (1),
\begin{equation}
(1\mid 2) \equiv \frac{1- \cos \theta_{p_1p_2}}
{(1- \cos \theta_{kp_1})(1- \cos \theta_{kp_2})}
\end{equation}
is positive definite.
It is singular both
as $\theta_{kp_1} \rightarrow 0$ and as $\theta_{kp_2} \rightarrow 0$.
We can write it as the sum of two terms
each of which is singular in one direction only.
\begin{equation}
(1\mid 2)= \frac{1}{2} \left\{ (1\mid 2)+
\frac{1}{1- \cos \theta_{kp_1}}- \frac{1}{1- \cos \theta_{kp_2}}
\right\} +\left\{1 \leftrightarrow 2\right\}
\end{equation}
In the first parenthesis on the rhs of (3),
the singularity in $(1\mid 2)$ as $\theta_{kp_2}
\rightarrow 0$ is cancelled
by the third term.
When we integrate it over the azimuthal angle
$\phi$ (around the direction
of $\vec p_1$),
making use of the relation
\begin{equation}
\cos \theta_{kp_2} =
\cos \theta_{kp_1} \cos \theta_{p_1p_2} + \sin \theta_{kp_1}
\sin \theta_{p_1p_2}\cos \phi \;,
\end{equation}
we find that it vanishes outside a cone $\theta_{kp_1}>\theta_{p_1p_2}$:
\begin{equation}
\frac{1}{2\pi } \int_0^{2\pi} \frac{1}{2}
\left\{ (1\mid 2)+ \frac{1}{1- \cos \theta_{kp_1}}-
\frac{1}{1- \cos \theta_{kp_2}} \right\} \d\phi
= \frac{1}{1- \cos \theta_{kp_1}}
\Theta (\theta_{p_1p_2} -\theta_{kp_1})\;,
\end{equation}
where $\Theta(x)$ is the step function: $\Theta(x)=0$ for
$x\leq0$ and $\Theta(x)=1$ for $x>0$.

This result is the simplest example of the exact AO.
The AO is called \lq\lq exact" in the following sense:
With the angular restriction $\theta_{kp_1} <\theta_{p_1p_2}$,
we can simply use $1/(1-\cos\theta_{kp_1})$
for the emission amplitude.
Although the exact amplitude before the azimuthal integration
is a more complicated quantity,
which is identical to $1/(1-\cos\theta_{kp_1})$
only at a very small angle,
the approximation is exact in evaluating an angle-averaged quantity.

\subsection{Soft-Gluon Emission from a Gluon}

Let us next suppose that an energetic quark emits a gluon at a
momentum $k_1$ at a small angle $\theta_{k_1p_1} \ll 1$.
Two diagrams contribute to the soft gluon emission
(at a momentum $k_2$) from them (Fig.2),
if the invariant mass $k_2\cdot p_1$ is
much smaller than $k_1\cdot p_1$.
The factorised amplitude in the eikonal approximation is
\begin{equation}
gf^{abc} i\tau_c \frac{k_1^{\mu}}{k_2 \cdot k_1}
-g\tau_a \tau_b \frac{p_1^{\mu}}{k_2 \cdot p_1}
\end{equation}
where $p_1$ is the momentum of the quark, and
$f^{abc}$ is the structure constant of SU(3):
\begin{equation}
[\tau^a,\tau^b]=if^{abc}\tau_c
\end{equation}
When the emission angle of the soft gluon $\theta_{k_2p_1}$ is much
larger than $\theta_{k_1p_1}$,
we have $\theta_{k_2k_1}\simeq\theta_{k_2p_1}$, and
\begin{equation}
\frac{k_1^{\mu}}{k_2 \cdot k_1}
=\frac{k_1^{\mu}/k_1}{k_2(1-\cos \theta_{k_2k_1})}
\simeq \frac{p_1^{\mu}}{k_2 \cdot p_1}
=\frac{p_1^{\mu}/p_1}{k_2(1-\cos \theta_{k_2p_1})}
\end{equation}
Making use of the commutation relation
of the colour matrix (7), we obtain
\begin{equation}
g\left( if^{abc} \tau_c \frac{k_1^{\mu}}{k_2 \cdot k_1}-\tau_a \tau_b
\frac{p_1^{\mu}}{k_2 \cdot p_1}\right)
\simeq -g\tau_b \tau_a \frac{p_1^{\mu}}{k_2 \cdot p_1}
\end{equation}
The colour structure on the rhs of (9) implies that the
large-angle soft gluon emission factorises as though it were
emitted before the quark emits a parallel gluon.
In general, a large-angle soft gluon is emitted coherently from
a narrow jet:
the gluon cannot resolve the polarisation in the jet, and is emitted
with its net charge.
Consequently, the secondary soft-gluon emission from the gluons
emitted inside a narrow jet can occur only at a small angle.

Let us now analyse the coherence and the AO in the same process
on the cross section level.
The soft-gluon emission amplitude factorises
from the $e^+e^- \to q\bar{q}g$ cross section:
\begin{eqnarray}
\d\sigma (e^+e^- \to q\bar{q}gg)&\simeq& \d\sigma (e^+e^- \to q\bar{q}g)
\frac{4\pi\alpha_s\d^3 k_2}{(2\pi )^32k_2}
\left[C_F\frac{2p_1 \cdot p_2}{(k_2 \cdot p_1)(k_2 \cdot p_2)} \right.
\nonumber\\
\nonumber\\
& & +\left. \frac{C_A}{2} \left\{
\frac{2k_1 \cdot p_1}{(k_2 \cdot k_1)(k_2 \cdot p_1)}
+ \frac{2k_1 \cdot p_2}{(k_2 \cdot k_1)(k_2 \cdot p_2)}
- \frac{2p_1 \cdot p_2}{(k_2 \cdot p_1)(k_2 \cdot p_2)} \right\} \right]
\nonumber\\
\nonumber\\
&=&\d\sigma (e^+e^- \to q\bar{q}g) \frac{\d^3 k_2}{2\pi^2k_2}
\nonumber\\
\nonumber\\
& &\times
\left[ C_F\alpha_s(p_1\mid p_2)
+\frac{C_A}{2}\alpha_s\left\{(k_1\mid p_1)+(k_1\mid p_2)-
(p_1\mid p_2)\right\}\right]\;,
\end{eqnarray}
where
\begin{eqnarray}
(k_1\mid p_1) &=& \frac{1- \cos \theta_{k_1p_1}}
{(1- \cos \theta_{k_2k_1})(1- \cos \theta_{k_2p_1})}\nonumber\\
(k_1\mid p_2) &=& \frac{1- \cos \theta_{k_1p_2}}
{(1- \cos \theta_{k_2k_1})(1- \cos \theta_{k_2p_2})}\nonumber\\
(p_1\mid p_2) &=& \frac{1- \cos \theta_{p_1p_2}}
{(1- \cos \theta_{k_2p_1})(1- \cos \theta_{k_2p_2})}\;.
\end{eqnarray}
The gluon colour charge $C_A$ is defined by
$\displaystyle \sum_{a,b} f^{abc} f^{abc'} = C_A \delta^{cc'}\, (C_A =3$
for SU(3)).
The first term, proportional to $C_F$, represents the emission
from the $q\bar{q}$-pair.
The contribution proportional to $C_A$ consists of three terms :
\begin{equation}
\frac{C_A\alpha_s}{2}\left\{(k_1\mid p_1)+(k_1\mid p_2)
-(p_1\mid p_2)\right\}\;,
\end{equation}
each of which has an angular form similar
to the emission from the $q\bar{q}$.
In (12), the singularity in the direction of the quark ($\vec{p}_1$)
is cancelled between the first and the third terms,
while the singularity in the direction of the antiquark ($\vec{p}_2$)
is cancelled between the second and the third term.
Therefore, the angular factor (12) has the singularity only in the
direction of the gluon ($\vec{k}_1$).
It thus corresponds to the emission from the $k_1$-gluon
(mainly directed near the direction of $\vec k_1$).

Let us now assume that the $k_1$-gluon is emitted
at a small angle from the quark
($\theta_{k_1p_1} \ll \theta_{p_1p_2} \simeq \theta_{k_1p_2}$).
As we have seen above, the soft-gluon amplitude from a gluon
vanishes when the emission angle of the soft gluon from the $k_1$-gluon
$\theta_{k_2k_1}$ is much larger than $\theta_{k_1p_1}$.
We, therefore, are interested in the case
that the soft-gluon emission angle $\theta_{k_2k_1}$
is comparable to or smaller than $\theta_{k_1p_1}$.
It implies that the angle between the soft gluon and the antiquark
$\theta_{k_2p_2}$ is nearly equal to
$\theta_{p_1p_2}(\simeq\theta_{k_1p_2}$).
Consequently, (12) can be approximated as
\begin{equation}
(12) \simeq \frac{C_A\alpha_s}{2}\left\{(k_1\mid p_1)
+\frac{1}{1-\cos \theta_{k_2k_1}}
-\frac{1}{1-\cos \theta_{k_2p_1}}\right\}\;.
\end{equation}
After the azimuthal angle averaging, we obtain
\begin{equation}
(13) \longrightarrow \frac{C_A\alpha_s}{1-\cos \theta_{k_1k_2}}\,
\Theta(\theta_{k_1p_1}-\theta_{k_2k_1})
\end{equation}
Thus the AO is exact at small angles.

The result obtained above can be generalised to higher orders.
Let us consider successive soft-gluon emission
with strongly ordered soft momenta $p_1\gg k_1\gg k_2\gg k_3\gg\cdots$.
We construct Feynman diagrams by successive soft-gluon insertion
into less soft lines.
At each step, the factorised soft-gluon emission
amplitude (at a momentum $k_{i+1}$) can be divided
into two types of contribution.
One of them is the soft-gluon emission from the colour configuration
under which the previous $i$-th gluon (at a momentum $k_i$) was emitted.
The first term (proportional to $C_F$) in the factorised amplitude
of the $k_2$-gluon emission in (10) is its simplest example.
This type of contribution takes exactly the same form as the
$i$-th gluon emission.

On the other hand, the new colour configuration
caused by the emission of the $i$-th gluon
gives rise to a new contribution.
Suppose that the emission angle of the $i$-th gluon
from the ($i-1$)-th gluon is small.
There are three types of contribution
which are relevant to this process.
One is from the interference term between the emission from the
$i$-th gluon and the $(i-1)$-th gluon
(for $i=1$, it is the first term in (12)).
The second type is from the interference between the $i$-the gluon and
the sum of all the possible soft-gluon insertion to the final
particles except the $i$-th and $(i-1)$-th gluons.
For $i=1$, it is the second term in (12).
For $i\geq 2$, the soft-gluon insertion
can be done on more than one lines
(not just on the $p_2$-line in the second term of (12)).
Note that the angular factor is insensitive to the direction
of the momentum of the line on which the soft gluon is inserted,
as is the case in the second term on the rhs of (13).
With the help of the Ward-Takahashi identity\cite{war},
we can easily prove
that the sum of all the possible insertion
gives the same angular factor
as the insertion to a single (non-soft) parton which forms a colour
singlet with the $i$-th and $(i-1)$-th gluons.

The third type is from the interference between the $(i-1)$-th gluon
and the sum of all the possible soft-gluon insertion to the final
particles except the $i$-th and $(i-1)$-th gluons.
The contribution from this type
of interference includes the independent
emission from $(i-1)$-th gluon mentioned earlier.
We, however, have an additional contribution, associated with the
non-commutativity of the non-Abelian gauge coupling.
The latter can be analysed in a similar way as the second type
(for $i=1$, it is the third term in (12)).

We thus obtain an angular factor
\begin{equation}
C_A\alpha_s\left\{(k_{i-1}\mid k_i)+\frac{1}
{1-\cos \theta_{k_{i+1}k_i}}
-\frac{1}{1-\cos \theta_{k_{i+1}k_{i-1}}}\right\}\;.
\end{equation}
The azimuthal averaging of (15) leads to the exact AO form
\begin{equation}
(15)\,\longrightarrow\,\frac{C_A\alpha_s}{1-\cos \theta_{k_{i+1}k_i}}\,
\Theta (\theta_{k_ik_{i-1}}-\theta_{k_{i+1}k_i})
\end{equation}

The picture obtained above can be summarised as follows\cite{tes}:
The initial $q\bar q$-pair emits a number of gluons
independently (the primary gluons).
Each primary gluon emits softer secondary gluons
independently of one another
at angles smaller than the emission angle of the primary gluon.
Each of the secondary gluons emits
softer gluons independently of one another
at angles smaller than the emission angle
of the secondary gluon, and so on.

In this article, we are interested in the 1PI cross section.
If we integrate it over the registered momentum and divide it
by the total cross section,
we obtain the total particle multiplicity per event.
In the soft-gluon emission described above,
the parton cascade ramifies into many branches.
Let us take the branch to which the registered particle belongs.
The emission angles along this branch is in the decreasing order,
$\theta_{p_1p_2}>\theta_{k_1p_1}>\theta_{k_2k_1}>
\theta_{k_3k_2}\cdots$.
There are other branches corresponding to the independent emission from
the parent partons.
The integration over the momentum of the unregistered particles gives
rise to the logarithmic divergence.
Factorisation takes place for each soft-gluon emission amplitude
(either virtual or real),
and the cancellation of the singularity
between real emission of the unregistered particle
and virtual emission takes place,
owing to the KLN theorem\cite{kin}.
There are no residual large logarithms after the cancellation.
We therefore can neglect the independent emission in the evaluation
of the 1PI cross section at the next-to-leading order.

If we go back to the idea of the coherence in the soft-gluon emission
mentioned earlier in this subsection,
we obtain an alternative picture of the
angle-ordered branching process:
The largest-angle soft gluon from the quark (similarly from the
antiquark) is first emitted at an angle smaller than the angle between
the $q\bar{q}$-pair and with the colour charge of a quark
(the net colour of the jet), because of the coherence.
Then from the second-largest-angle soft gluon is emitted from
the quark in a similar way, and so on.
These gluons are angle-ordered, but not energy-ordered.
(As a matter of fact, the largest-angle gluon tends to be
softer than others\cite{mar}, because of the
transverse momentum cutoff in the emission.)

On the other hand, the emitted softer gluons emit softer gluons.
The emission angles of the latter gluons do not exceed the emission
angle of the gluon from which they are emitted, and so on.
Along this side of branching,
gluons are both angle-ordered and energy-ordered.

It is on the latter picture that the
coherent shower Monte Carlo simulation
of the QCD multiparticle production is based\cite{mar}\cite{gus}.
On the cross section level, the two pictures are indeed equivalent.
The \lq\lq independent" emission in the former picture can be
angularly ordered, just with dropping the Bose-statistical
factor $1/n!$, where $n$ is the number of soft gluons emitted
independently from the parent parton.

\section{Modification due to the Azimuthal Averaging}

As is described in the previous section,
the AO is exact at the next-to-leading order
for angle-averaged quantities,
such as the total multiplicity and the particle energy spectrum.
The AO, however, is by no means obvious
for the angular distribution of particles.

In order to illustrate how the azimuth-averaged amplitude
(AO with a step function) differs from the exact amplitude,
let us examine the simplest case\cite{tes}: One soft-gluon
emission amplitude from an energetic quark-antiquark ($q\bar q$) pair.

In fact, the order among angles is not a Lorentz invariant quantity.
Therefore, the distribution of the soft gluons generated with the AO
depends on the Lorentz frame in which the angles are defined.
Suppose, for example,
that the initial $q\bar q$ pair is produced at the rectangle
from a gauge-invariant vertex (e.g. $\!$the electromagnetic current).
According to the AO, the soft-gluon radiation angle
from the quark has to be
less than $\pi /2$:
namely in the quark-side hemisphere divided by the plane
perpendicular to the quark direction.
Similarly, the soft-gluon radiation is in the antiquark-side hemisphere
divided by the plane perpendicular to the antiquark direction.
We, therefore, have four solid-angle regions divided by
two planes (Fig.3).
In the region I in Fig.3, we have gluon radiation only from the quark,
while in the region II we have radiation only from the antiquark.
In the region III, the gluon can come
both from the quark and from the antiquark.
Finally, in the region IV, we see no soft-gluon radiation.

Now let us boost these angular regions to the centre-of-mass system of
the $q\bar q$ (we neglect the recoil from the soft-gluon radiation).
We then find an angular region where no gluons are emitted (around the
direction of the boost).
Such a region does not exist
in the original soft-gluon emission amplitude (1),
which is azimuthal-symmetric in the c.m. $\!$frame.

This shortcoming of the AO in the angular distribution
may be avoided at $O(\alpha_s)$ if we define it in the centre-of-mass
frame of $q\bar q$.
In this case, the azimuthal integration is trivial.
The division of the amplitude into two parts is in this frame
\begin{equation}
(1\mid 2) = \frac{2}{1-\cos^2\theta_{kp_1}}=
\frac{1}{1-\cos\theta_{kp_1}}+\frac{1}{1-\cos\theta_{kp_2}}\;.
\end{equation}
Indeed, there is no larger angle than $\pi$, and the the AO
does not restrict the emission angle.

Whenever the AO is non-trivial, however,
the azimuth-averaged expression (with a step function)
is not identical to the exact cross section.
In the following, we examine if there is any next-to-leading order
correction to the exact AO in the angular distribution of particles.
The quantity is obtained
by integrating the 1PI cross section
over the registered energy.

Let us first analyse the two soft-gluon emission
amplitude at the tree level
in the c.m. $\!$frame.
The cross section is given by
\begin{eqnarray}
\d \sigma_{\rm tree}
&=& \sigma_0 \, C_F \alpha_s \frac{2p_1 \cdot p_2}
{(k_1 \cdot p_1)(k_1 \cdot p_2)} \left[ C_F \alpha_s
\frac{2p_1 \cdot p_2}{(k_2 \cdot p_1)(k_2 \cdot p_2)} \right.
\nonumber\\
\nonumber\\
& & \left. + \frac{C_A \alpha_s}{2}
\left\{ \frac{2k_1 \cdot p_1}{(k_2 \cdot k_1)(k_2 \cdot p_1)}
+\frac{2k_1 \cdot p_2}{(k_2 \cdot k_1)(k_2 \cdot p_2)}
-\frac{2p_1 \cdot p_2}{(k_2 \cdot p_1)(k_2 \cdot p_2)} \right\} \right]
\nonumber\\
\nonumber\\
& & \times \frac{\d ^3 k_1}{(2\pi)^2 k_1} \frac{\d ^3 k_2}{(2\pi)^2 k_2}
\end{eqnarray}
where we have assumed $p_1,p_2 \gg k_1 \gg k_2$ (see (10)).

Suppose that $k_2$ is the registered momentum.
We integrate (18) over $k_1$ to obtain the 1PI cross section.
Note that (18) is symmetric under the exchange $k_1
\leftrightarrow k_2$.
Therefore, it can be written as
\begin{eqnarray}
\d \sigma_{\rm tree}
&=& \sigma_0 \, C_F \alpha_s \frac{2p_1 \cdot p_2}{(k_2 \cdot p_1)
(k_2 \cdot p_2)} \left[ C_F \alpha_s
\frac{2p_1 \cdot p_2}{(k_1 \cdot p_1)(k_1 \cdot p_2)} \right.
\nonumber\\
\nonumber\\
& & \left. + \frac{C_A \alpha_s}{2} \left\{
\frac{2k_2 \cdot p_1}{(k_1 \cdot k_2)(k_1 \cdot p_1)}
+\frac{2k_2 \cdot p_2}{(k_1 \cdot k_2)(k_1 \cdot p_2)}
-\frac{2p_1 \cdot p_2}{(k_1 \cdot p_1)(k_1 \cdot p_2)} \right\} \right]
\nonumber\\
\nonumber\\
& & \times \frac{\d ^3 k_2}{(2\pi)^2 k_2} \frac{\d ^3 k_1}{(2\pi)^2 k_1}
\end{eqnarray}
We may use (19) both for $k_1>k_2$ and for $k_2>k_1$.

When integrated over $k_1$, the first term in the curly bracket
on the rhs of (19), proportional to the quark colour charge $C_F$,
gives a logarithmic divergence, which is cancelled by the
virtual correction at the vertex of the $q \bar q$-pair production.
Similarly, the integration of the second term (proportional to the
gluon colour charge $C_A$) over $k_1$ with $k_1<k_2$ gives
a logarithmic singularity, which is cancelled by the virtual correction
at the $k_2$-gluon emission vertex.
No large logarithms remain after the cancellation.
Because we are interested in the terms with large double-logarithms,
we consider only the second term with $k_1>k_2$.

If the registered angle $\theta =\theta_{k_2p_1}$ is small,
we can make use of the approximation:
\begin{eqnarray}
& &\frac{1-\cos\theta}
{(1-\cos\theta_{k_1k_2})(1-\cos\theta_{k_1p_1})}
+\frac{1-\cos\theta_{k_2p_2}}
{(1-\cos\theta_{k_1k_2})(1-\cos\theta_{k_1p_2})}
\nonumber\\
\nonumber\\
& &-\frac{1-\cos\theta_{p_1p_2}}
{(1-\cos\theta_{k_1p_1})(1-\cos\theta_{k_1p_2})}
\nonumber\\
\nonumber\\
&\simeq &
\frac{1-\cos\theta}
{(1-\cos\theta_{k_1k_2})(1-\cos\theta_{k_1p_1})}
+\frac{1}{1-\cos\theta_{k_1k_2}}-\frac{1}{1-\cos\theta_{k_1p_1}}
\end{eqnarray}
Its azimuthal integration over $\vec k_1$ gives
the step function $\Theta (\theta -\theta_{k_1k_2})$,
similarly to (14).
Namely, the exact AO takes place also for the opposite energy ordering
(we call it the \lq\lq reverse exact AO"),
because of the symmetry between (18) and (19).
We thus obtain
\begin{eqnarray}
\frac{\d \sigma}{\d (\cos\theta )\,\d k_2}
&=& \sigma_0 \frac{C_F \alpha_s}{\pi\,k_2}
\frac{C_A \alpha_s}{\pi(1-\cos\theta)}
\int^{p_1}_{k_2}\frac{\d k_1}{k_1}\int^1_{\cos\theta}
\frac{\d (\cos\theta_{k_1k_2})}{1-\cos\theta_{k_1k_2}}
\nonumber\\
\nonumber\\
&=& \sigma_0 \frac{C_F \alpha_s}{\pi\,k_2}
\frac{C_A \alpha_s}{\pi(1-\cos\theta)} \ln \frac{ p_1 }{k_2}
\int^1_{\cos\theta}
\frac{\d (\cos\theta_{k_1k_2})}{1-\cos\theta_{k_1k_2}}\;.
\end{eqnarray}

The expression (20) has the form of the product of two amplitudes:
the one-gluon emission amplitude (primary gluon)
at the registered angle $\theta$,
and the soft-gluon emission amplitude
(from the primary gluon) integrated over
the phase space within the angle-ordered cone.
The latter is identical to the gluon multiplicity (at $O(\alpha_s)$)
in the subjet of the opening angle identical to $\theta$.

In the next section,
where we evaluate the angular distribution to all orders in $\alpha_s$,
we shall make systematical use of
the identification of the emission angle of the primary gluon
with the registered angle $\theta$.
The reverse exact AO above implies that at small angles
there is no next-to-leading order correction to this approximation.

\section{Resummation of Double-Logarithms}

Because the integration over the momentum of each unregistered gluon
gives rise to a double-logarithm,
higher order contributions
in the perturbative series in $\alpha_s$
are not negligible in the high-energy limit
compared with the contributions at the lowest orders.
We, therefore, have to sum the large double-logarithms up to all orders
in $\alpha_s$ to redefine the perturbative expansion (resummation).
The resummation of the leading double logarithm is done by
the double-log approximation\footnote{
This approximation, including its single-logarithmic
next-to-leading order corrections, is given as a modified version
of the more conventional leading-log approximation\cite{gri};
hence the name \lq\lq modified leading-log approximation" (MLLA).}
(DLA)\cite{dla}\cite{mun}.

When many gluons are emitted (with a high power in $\alpha_s$),
the matrix element shows, in general,
complicated dependence on their angular configuration.
The amplitude, however, takes particularly simple form
for the configuration in which we are interested in DLA.
The largest logarithm at each order in $\alpha_s$
comes from the phase space configuration
in which all the momenta are {\it strongly} ordered:
\begin{equation}
p_1\gg k_1 \gg k_2 \gg k_3 \gg \cdots \gg q
\end{equation}
where $p_1$ is the energy of the initial quark,
$k_1$ is that of the primary
soft gluon, $k_2$ is that of softer secondary gluon, etc., and
$q$ is the energy of the registered particle.
The emission angles are also strongly ordered.
Because of the angular ordering (or coherence), we have
\begin{equation}
\theta_{k_1p_1} \gg \theta_{k_2k_1} \gg
\theta_{k_3k_2} \gg \cdot \cdot \cdot
\end{equation}
where $\theta_{k_1p_1}$ is the emission angle of the primary soft gluon
from the direction of the quark, $\theta_{k_2k_1}$ is that of
the softer secondary gluon from the direction
of the primary gluon, etc.
If $\theta$ is the angle between the
registered particle and the energetic
quark, we have
\begin{equation}
\theta \simeq \theta_{k_1p_1}\;.
\end{equation}

The $\theta$-distribution is given at the leading order by
\begin{eqnarray}
\frac{\d <n>}{\d \theta}&=& \frac{C_F\alpha_s}{\pi^2}
\int^{p_1}\frac{\d k_1}{k_1}
\int^{\theta}\frac{\d \theta_1\d\phi_1}{\theta_1}
\nonumber\\
\nonumber\\
&\times&
\left\{\delta (\vec\theta-\vec\theta_1)
+ \sum_{n=1}\prod^n_{j=1}\left[\frac{C_A\alpha_s}{\pi^2}
\int^{k_j}\frac{\d k_{j+1}}{k_{j+1}} \int^{\theta_j}
\frac{\d \theta_{j+1}\d\phi_{j+1}}{\theta_{j+1}}\right]
\delta (\vec\theta-\sum_{i=0}^n\vec\theta_{i+1})\right\}
\nonumber\\
&\approx &
\frac{2C_F\alpha_s}{\pi\theta}\int^{p_1}
\frac{\d k}{k}M_g(k^2\theta^2)\;.
\;\;\;(\theta_1=\theta_{k_1p_1},\;\theta_i=\theta_{k_ik_{i-1}}\;
{\rm for}\;i\geq 2)
\end{eqnarray}
Use is made of (24) to obtain the last expression, where
$M_g(k^2\theta^2)$ is the multiplicity in a gluon jet with the
energy $k$ and the opening angle $\theta$:
\begin{equation}
M_g(k^2\theta^2)=1+\sum_{n=0}\prod^n_{j=0}\frac{2C_A\alpha_s}{\pi}
\int^{k_j}\frac{\d k_{j+1}}{k_{j+1}}
\int^{\theta_j}\frac{\d \theta_{j+1}}{\theta_{j+1}}
\;.\;\;\;(k_0=k,\,\theta_0=\theta)
\end{equation}

The series summation can be represented as an integral equation
\begin{equation}
M_g(k^2\theta^2)=\frac{2C_A\alpha_s}{\pi} \int^k\frac{\d k_1}{k_1}
\int^{\theta}\frac{\d \theta_1}{\theta_1}M_g(k_1^2\theta_1^2)+1
\end{equation}
At asymptotic high energies, the equation (27) can be approximated by
an homogeneous equation:
\begin{equation}
M_g(k^2\theta^2)\approx
\frac{2C_A\alpha_s}{\pi}\int^k\frac{\d k_1}{k_1}
\int^{\theta}\frac{\d \theta_1}{\theta_1}
M_g(k_1^2\theta_1^2)\;.
\end{equation}
The solution of the homogeneous equation (28) is given by
\begin{eqnarray}
M_g(k^2\theta^2)&=&C(k^2\theta^2)^{\gamma}
\nonumber\\
\gamma&=&\sqrt{\frac{C_A\alpha_s}{2\pi}}\;,
\end{eqnarray}
where the normalisation constant $C$ is not determined
by the homogeneous equation (28).

The next-to-leading order corrections to (25) come
from three cases\cite{dok2}\cite{dok1}\cite{mue2}:

(i) When the strong energy ordering (22) does not hold
at one of the branching points in the successive gluon emission:
\begin{equation}
\frac{k_i}{k_{i-1}}=O(1)\;.
\end{equation}
For the non-soft emission,
the integration kernel has to be modified in (25)-(28):
\begin{eqnarray}
\int_0^{p_1}\frac{\d k_1}{k_1} \; &\to& \;
\int_0^1\d x_1 \left( \frac{1}{x_1}-\frac{3}{4}\right)
\;\;,\;(x_1=\frac{k_1}{p_1})\;
({\rm for\,the\,emission\,from\,a\,quark)}
\nonumber\\
\nonumber\\
\int_0^{k_{i-1}}\frac{\d k_i}{k_i} \; &\to& \;
\int_0^1\d x_i \left( \frac{1}{x_1}-\frac{11}{12}-\frac{N_f}{6C_A}
-\frac{N_fC_F}{3C_A^2}\right)
\;,\;\;({\rm from\,the}\,k_{i-1}{\rm -gluon)}
\end{eqnarray}
where $N_f$ is the number of active quark flavours.

(ii) The virtual corrections at the gluon-emission vertex run the
effective QCD coupling.
Replacing $\alpha_s(k_{\perp}^2)$ ($k_{\perp}=k_1\sin\theta_1$)
for the QCD coupling on the rhs
of the integral equation (28) (together with the replacement (31)),
we obtain its asymptotic solution:
\begin{eqnarray}
M_g(k_{\perp}^2)&\approx& C\left[\ln (k_{\perp}^2/\Lambda_{QCD}^2)
\right]^{\gamma_1}
\exp \left[2\gamma_0\sqrt{\ln (k_{\perp}^2/
\Lambda_{QCD}^2)}\right]\;,\\
\nonumber\\
\gamma_0 &=& \sqrt{\frac{C_A}{2\pi b_0}}\;,
\nonumber\\
\gamma_1 &=& -\frac{1}{4}-\frac{N_f}{6\pi b_0}
\left(1-\frac{C_F}{C_A}\right)\;,
\nonumber\\
b_0 &=& \frac{11C_A-2N_f}{12\pi}\;.\nonumber
\end{eqnarray}

With the running coupling, the anomalous dimension $\gamma$ is
replaced by
\begin{equation}
\gamma \equiv \frac{\d\,\ln M_g(k_{\perp}^2)}{\d\,\ln k_{\perp}^2}=
\frac{\gamma_0}{\sqrt{\ln (k_{\perp}^2/\Lambda_{QCD}^2)}}
+\frac{\gamma_1}{\ln (k_{\perp}^2/\Lambda_{QCD}^2)}\;.
\end{equation}
The logarithmic integration of $M_g$ gives a factor $1/\gamma$,
in stead of a logarithm.

(iii) When one of the soft-gluon emission angle
is of the same order of magnitude as the preceding one.
If $\theta_{k_ik_{i-1}}/\theta_{k_{i-1}k_{i-2}}=O(1)$ $(i\geq 3)$
(but the strong AO holds between $\vec k_1$ and $\vec k_2$,
$\theta_{k_1p_1}\gg \theta_{K_2k_1}$),
then we can make use of the approximation (24),
and the azimuthal averaging over $\vec k_i$ (around $\vec k_{i-1}$)
leads to the exact AO discussed in sect.2.

The next-to-leading correction to the AO, therefore,
can occur only for the configuration:
\begin{equation}
\frac{ \theta_{k_2k_1} }{ \theta_{k_1p_1} }=O(1), \quad
\vec{\theta}_{k_1p_1}+\vec{\theta}_{k_2k_1}=
\vec{\theta}_{k_2p_1} \simeq \vec{\theta}\;.
\end{equation}
In this case, we have to use a more exact form
for the soft-gluon emission.

At a small angle ($\theta\ll 1$), however,
we can make use of the reverse exact AO discussed in the last section.
The angular particle density is given by multiplying the tree-level
two-gluon emission cross section (divided by the total cross section)
with the multiplicity in the $k_2$-subjet with an opening angle
identical to the emission angle of the $k_2$-gluon $\theta_{k_2k_1}$.
Following the steps which has led to (21), we obtain
\begin{eqnarray}
\frac{\d <n>}{\d \theta}
&=&
\frac{2C_F\alpha_s}{\pi}\frac{2C_A\alpha_s}{\pi}\frac{1}{\theta}
\int^{p_1}\frac{\d k_1}{k_1}\int^{k_1}\frac{\d k_2}{k_2}
\int^{\theta}\frac{\d \theta_{k_1k_2}}{\theta_{k_1k_2}}
M_g(k_2^2\theta_{k_1k_2}^2)
\nonumber\\
\nonumber\\
&=&
\frac{2C_F\alpha_s}{\pi}\frac{C_A\alpha_s}{2\pi}\frac{1}{\gamma^2}
\frac{1}{\theta}\int^{p_1}\frac{\d k_1}{k_1}M_g(k_1^2\theta^2)
\nonumber\\
\nonumber\\
&=&
\frac{2C_F}{C_A}\frac{\gamma}{\theta} M_g(p_1^2\theta^2)\;,
\end{eqnarray}
where use is made of the value of $\gamma$, (29),
to obtain the last expression,
which is identical to the leading order result
({\it cf.} the last expression in (25)).

Finally, let us discuss the coherent Monte Carlo simulation
which generates particles with the AO\cite{mar}\cite{gus}.
It appears that the particle flow generated by the simulation
is in sharp contrast with the picture of the reverse exact AO,
discussed in the previous section and used to prove (35) above.
In the simulation, the $k_1$-gluon emitted at a large angle
can emit a softer $k_2$-gluon at a similar large angle
to form a small angle
(i.e.$\,\vec\theta_{k_1p_1}+\vec\theta_{k_2k_1}=\vec\theta$,
\,$\theta_{k_1p_1}>\theta_{k_2k_1}\gg\theta$).
Such a process would not give any contribution
under the picture of the
reverse exact AO in (21).

In fact, when $\theta_{k_1p_1}\gg \theta$,
the partial amplitude (20) gives a positive contribution
for $\theta_{k_2k_1}<\theta_{k_1p_1}$,
while the contribution becomes negative
for $\theta_{k_2k_1}>\theta_{k_1p_1}$.
The contributions cancel out one another in average,
which results in the reverse exact AO in (21).
On the other hand, the Monte Carlo simulation assigns a positive
probability to the former case, and zero to the latter.
Thus the positive probability
for the large angle process is not cancelled.
This leads to the overestimation of the particle density
at the next-to-leading order in the forward direction.

The simulation, however, somewhat underestimates the contribution
from the small angle process
in which $\theta>\theta_{k_2k_1}>\theta/2$,
because of the AO.
In fact, after integrating over the direction of $\vec k_1$,
the errors in the simulation cancel out.
The exactness of the angular distribution at the next-to-leading order
is required at small angles by the fact that the small angle
region contributes to the total multiplicity at the leading order.
Because the AO is exact in the total multiplicity,
as was discussed in sect.2,
the particle density in the direction of a jet generated
from a massless quark in the simulation with the exact AO
cannot have next-to-leading order correction.

\section{Corrections at Large Angles}

Let us next analyse the correction to the AO at a large angle.
Because the large-angle particles contribute to the total multiplicity
at the next-to-leading order,
it is possible that the correction to the AO appears
at the next order at a large angle .

We start with the two-soft-gluon emission cross section
at the tree level (19).
We are interested in the term proportional
to the gluon colour charge $C_A$:
\begin{eqnarray}
\d \sigma(g\rightarrow g)
&=& \sigma_0 \, C_F \alpha_s
\frac{2p_1 \cdot p_2}{(k_2 \cdot p_1)(k_2 \cdot p_2)}
\frac{\d ^3 k_2}{(2\pi)^2 k_2}
\frac{\d ^3 k_1}{(2\pi)^2 k_1}
\nonumber\\
\nonumber\\
&\times & \frac{C_A\alpha_s}{2} \left\{
\frac{2k_2 \cdot p_1}{(k_1 \cdot k_2)(k_1 \cdot p_1)}
+ \frac{2k_2 \cdot p_2}{(k_1 \cdot k_2)(k_1 \cdot p_2)}
- \frac{2p_1 \cdot p_2}{(k_1 \cdot p_1)(k_1 \cdot p_2)} \right\}
\end{eqnarray}
As in sect.3, we fix $\theta=\theta_{k_2p_1}$,
and integrate over $k_1$ ($k_2<k_1<p_1$)
to obtain the 1PI cross section.

At a large angle ($\theta/\theta_{p_1p_2}=O(1)$), the condition
$\theta_{k_1k_2} \ll \theta_{k_2p_2} \simeq \theta_{k_1p_2}$ or
$\theta_{k_1p_1} \ll \theta_{p_1p_2} \simeq \theta_{k_1p_2}$
is not necessarily satisfied,
and we cannot make use of the approximation(20).
We therefore have to analyse each term in (36) more carefully.

The $k_1$-integration of each term can be performed with the use of the
decomposition similar to that in subsect.2.1.
The azimuthal integration of each term gives
\begin{eqnarray}
\frac{k_2 \cdot p_1 k_1^2}{(k_1 \cdot k_2)(k_1 \cdot p_1)}
&=& \frac{1}{2}\left\{
\frac{1}{(1-\cos\theta_{k_1k_2})(1-\cos\theta_{k_1p_1})}
+\frac{1}{1-\cos\theta_{k_1p_1}}-\frac{1}{1-\cos\theta_{k_1k_2}}
\right\}
\nonumber\\
&& +(p_1 \leftrightarrow k_2)
\nonumber\\
\nonumber\\
\longrightarrow
&\quad& \frac{\d(\cos\theta_{k_1p_1})}{k_1(1-\cos\theta_{k_1p_1})}
\Theta (\theta_{k_2p_1}- \theta_{k_1p_1})+(p_1 \leftrightarrow k_2),
\nonumber\\
\nonumber\\
\frac{k_2 \cdot p_2k_1^2}{(k_1 \cdot k_2)(k_1 \cdot p_2)}
&=& \frac{1}{2}\left\{
\frac{1}{(1-\cos\theta_{k_1k_2})(1-\cos\theta_{k_1p_2})}
+\frac{1}{1-\cos\theta_{k_1p_2}}-\frac{1}{1-\cos\theta_{k_1k_2}}
\right\}
\nonumber\\
&& +(p_2 \leftrightarrow k_2)
\nonumber\\
\nonumber\\
\longrightarrow
&\quad& \frac{\d(\cos\theta_{k_1p_2})}{k_1(1-\cos\theta_{k_1p_2})}
\Theta (\theta_{k_2p_2}- \theta_{k_1p_2})
+(p_2 \leftrightarrow k_2),
\nonumber\\
\nonumber\\
-\frac{p_1 \cdot p_2k_1^2}{(k_1 \cdot p_1)(k_1 \cdot p_2)}
&=& -\frac{1}{2}\left\{
\frac{1}{(1-\cos\theta_{k_1p_1})(1-\cos\theta_{k_1p_2})}
+\frac{1}{1-\cos\theta_{k_1p_1}}-\frac{1}{1-\cos\theta_{k_1p_2}}
\right\}
\nonumber\\
&& +(p_1 \leftrightarrow p_2)
\frac{1}{2}\left\{
\frac{p_1 \cdot p_2}{(k_1 \cdot p_1)(k_1 \cdot p_2)}+
\frac{1}{k_1\cdot p_1}-\frac{1}{k_1 \cdot p_2} \right\}
+(p_1 \leftrightarrow p_2)
\nonumber\\
\nonumber\\
\longrightarrow
&\quad& -\frac{\d(\cos\theta_{k_1p_1})}{k_1(1-\cos\theta_{k_1p_1})}
\Theta (\theta_{p_1 p_2}-\theta_{k_1 p_1}) - (p_1 \leftrightarrow p_2).
\end{eqnarray}
We thus obtain
\begin{eqnarray}
& &
\frac{C_A\alpha_s\d^3k_1}{(2\pi)^2 k_1} \left\{
\frac{2k_2 \cdot p_1}{(k_1 \cdot k_2)(k_1 \cdot p_1)}
+\frac{2k_2 \cdot p_2}{(k_1 \cdot k_2)(k_1 \cdot p_2)}
-\frac{2p_1 \cdot p_2}{(k_1 \cdot p_1)(k_1 \cdot p_2)} \right\}
\nonumber\\
\nonumber\\
&\simeq &
\frac{1}\pi \frac{C_A\alpha_s\d k_1}{k_1} \left\{
\int\frac{\d (\cos\theta_{p_1 k_1})}{1-\cos \theta_{p_1 k_1}}
\left[\Theta (\theta - \theta_{k_1p_1})
-\Theta(\theta_{p_1 p_2}-\theta_{k_1p_1})\right]
\right.
\nonumber\\
\nonumber\\
& &
+\,\int\frac{\d (\cos\theta_{k_1p_2})}{1-\cos \theta_{k_1p_2}}
\left[\Theta(\theta_{k_2p_2}-\theta_{k_1p_2})
-\Theta(\theta_{p_1 p_2}-\theta_{k_1p_2}) \right]
\nonumber\\
\nonumber\\
& & \left.
+\,\int\frac{\d (\cos\theta_{k_1 k_2})}{1-\cos \theta_{k_1 k_2}}
\left[ \Theta (\theta-\theta_{k_1k_2})+
\Theta (\theta_{k_2p_2}-\theta_{k_1k_2})\right]\right\}
\end{eqnarray}

Note that only the last integral (over $\cos\theta_{k_1k_2}$)
on the rhs of (38) gives logarithmic divergence.
Let us isolate the divergent part in the form of the exact AO
\begin{equation}
2\int\frac{\d (\cos\theta_{k_1k_2})}{1-\cos\theta_{k_1k_2}}
\Theta(\theta-\theta_{k_1k_2})\;.
\end{equation}
The rest gives its correction:
\begin{eqnarray}
&&-\,\int_{\cos\theta_{p_1 p_2}}^{\cos\theta}
\frac{\d (\cos\theta_{k_1 p_1})}{1-\cos \theta_{k_1 p_1}}
-\,\int_{\cos\theta_{p_1 p_2}}^{\cos\theta_{k_2p_2}}
\frac{\d (\cos\theta_{k_1p_2})}{1-\cos \theta_{k_1p_2}}
+\,\int_{\cos\theta_{k_2p_2}}^{\cos\theta}
\frac{\d (\cos\theta_{k_1 k_2})}{1-\cos \theta_{k_1 k_2}}
\nonumber\\
\nonumber\\
&=&
-\,\ln\frac{1-\cos\theta_{p_1 p_2}}{1-\cos\theta}
-\ln\frac{1-\cos\theta_{p_1 p_2}}{1-\cos\theta_{k_2p_2}}
+\ln\frac{1-\cos\theta_{k_2p_2}}{1-\cos\theta}
\nonumber\\
\nonumber\\
&=&
-2\ln\frac{1-\cos\theta_{p_1 p_2}}{1-\cos\theta_{k_2p_2}}\;.
\end{eqnarray}
Particularly, in the centre-of-mass frame of the $q\bar q$-pair
($\theta_{p_1p_2}=\pi,\,\theta_{k_2p_2}=\pi -\theta$), (40) is
\begin{equation}
-2\ln\frac{2}{1-\cos\theta_{k_2p_2}}=
-2\ln\frac{2}{1+\cos\theta} =
2\ln\left(\cos^2\frac{\theta}{2}\right)\;.
\end{equation}
Note that (41) vanishes as $\theta\rightarrow 0$.

Now let us generalise the above result to all orders.
At the next-to-leading order, we can make use of (34):
We identify the direction of the registered particle
with that of $\vec k_2$, and multiply the multiplicity of
the $k_2$-subjet to the $k_2$-gluon emission cross section.
The opening angle of the $k_2$-subjet is identified with
$\theta_{k_2k_1}$, owing to the AO.

We calculate the correction to the AO with the use
of solution of the fixed coupling integral equation (29).
In the result, we can replace the solution with running coupling (32)
for $M_g$, and the value of $\gamma$ (33) for the anomalous dimension.
Indeed, we can calculate the correction using running coupling
from the start without any difficulties.
Because the difference appears only at the next-to-next order,
we use the fixed coupling for the sake of notational simplicity.

When we discussed the resummation in the previous section,
all emission angles are assumed small.
If they are not necessarily small,
we have to be careful of the choice of the variables.
The emission angle, if it is large,
is not invariant under the boost in the direction of the parent parton.
In fact, the resummation is organised
in terms of the transverse momentum relative to the parent parton.
The multiplicity of the $k_2$-subjet
(with the $k_2$-gluon emitted from the $k_1$-gluon) is given by
\begin{equation}
M_g = (k_2^2 \sin^2\theta_{k_1 k_2})^{\gamma}
\end{equation}
where $k_2\sin\theta_{k_1 k_2}$ is the transverse momentum
relative to $\vec k_1$.
In (42) we have put unity for the normalisation constant $C$ in (29)
for notational simplicity.
With the AO part of the integration kernel (39), we obtain
\begin{eqnarray}
& &2\int\frac{\d\cos\theta_{k_1k_2}}{1-\cos\theta_{k_1k_2}}
M_g(k_2^2 \sin^2\theta_{k_1 k_2})
\Theta(\theta-\theta_{k_1k_2})
\nonumber\\
\nonumber\\
& &=\frac{2}{\gamma}
\left( 4k_2^2 \sin^2 \frac{\theta}{2} \right)^{\gamma}
=\frac{2}{\gamma}
\left( 2k_2^2 (1-\cos\theta) \right)^{\gamma}\;.
\end{eqnarray}

This result is not written in terms of the transverse momentum
$k_{2\perp}=k_2\sin\theta$, and not invariant under
the boost in the jet direction.

Let us now add lhs of (40) to the integration kernel
on the lhs of (43).
In the correction, the integration kernel over the angular variable
does not become singular in the integration range.
It implies that the angular variable in the argument
of the multiplicity
function may be set equal to $\theta\,(=O(1))$.
(We may put $\theta_{k_2p_2}$ or $\theta_{k_1k_2}$
in place of $\theta$.
In the absence of the singularity in the kernel
in the integration range,
the difference in the result is at higher orders.)
\begin{eqnarray}
& &-\int_{\cos\theta_{p_1p_2}}^{\cos\theta}
\frac{\d (\cos\theta_{k_1p_1})}{1-\cos \theta_{k_1p_1}}
M_g(k_2^2\sin^2\theta)
-\int_{\cos\theta_{p_1 p_2}}^{\cos\theta_{k_2p_2}}
\frac{\d (\cos\theta_{k_1p_2})}{1-\cos \theta_{k_1p_2}}
M_g(k_2^2\sin^2\theta)
\nonumber\\
\nonumber\\
& &
\qquad+\int_{\cos\theta_{k_2p_2}}^{\cos\theta}
\frac{\d \cos\theta_{k_1 k_2}}{1-\cos \theta_{k_1 k_2}}
M_g(k_2^2\sin^2\theta)
\nonumber\\
\nonumber\\
& &=
2\ln\left(\cos^2\frac{\theta}{2}\right) (k_2^2\sin^2\theta)^{\gamma}
+{\rm higher\,orders}\;.
\end{eqnarray}

Adding the correction (44) to (43), we obtain
\begin{eqnarray}
& &\frac{2}{\gamma}
\left( 4k_2^2 \sin^2 \frac{\theta}{2} \right)^{\gamma}
+2\ln\left(\cos^2\frac{\theta}{2}\right) (k_2^2\sin^2\theta)^{\gamma}
\nonumber\\
& &=
\frac{2}{\gamma}(k_2^2\sin^2\theta)^{\gamma}
+{\rm higher\,orders}\;.
\end{eqnarray}
The result is now boost invariant.

The total multiplicity is given by integrating (45) over the
energy variables $k_2$, $k_1$, and the angle variable $\theta$.
The $\theta$ integration of the AO part, the rhs of (43),
with the integration kernel $1/(1-\cos\theta)$ gives
$(2/\gamma^2)(2k_2^2)^\gamma$.
The similar integration of the correction, the rhs of (44)
over $\theta$, on the other hand, gives
\begin{eqnarray}
& &-\int_{-1}^1\frac{\d(\cos\theta)}{1-\cos\theta}
2\ln\frac{2}{1+\cos\theta}(k_2^2\sin^2\theta)^{\gamma}
\nonumber\\
\nonumber\\
& &=-2\zeta (2)(k_2^2)^{\gamma}
+{\rm higher\,orders}\;,
\end{eqnarray}
where
\begin{equation}
\zeta (2)=\int_0^1\frac{\d z}{1-z}\ln \frac{1}{z}=\frac{\pi^2}{6}\;.
\end{equation}
The correction, therefore, reduces the total multiplicity
obtained with the AO
by a factor $(1-\gamma^2\pi^2/6)$.
This result agrees with the dipole correction calculated by
Dokshitzer {\em et al.}\cite{dok2}\cite{dok1} using the planar gauge.

\section{Conclusion}

In this article, we have examined the next-to-leading order corrections
to the AO in the angular distribution of particles in a jet.
At the leading order, the angle of the registered particle is
identified with the direction of the subjet to which it belongs:
i.e. $\!$that of the primary gluon, which is directly
emitted from the initial energetic parton.
At a small angle (i.e. $\!$ near the jet direction),
we found that there is no next-to-leading corrections
to the AO.

In the particle flow generated by the coherent Monte Carlo simulation,
the large-angle emission of a primary gluon contributes
to the forward multiplicity, while such contribution is absent in the
exact cross section.
On the other hand, the contribution from the
small-angle emission of the primary gluon is somewhat reduced,
and the angular distribution obtained in this simulation is correct
at the next-to-leading order.

At a large angle, the correction to the AO
reduces the multiplicity by a factor of
$(1+\gamma\ln \,\cos^2(\theta/2))$.
The correction restores the boost invariance
of the angular distribution,
which is absent in the AO formalism.
In another word, the dipole correction appears only because
we took the angle variables to formulate the coherence.
There is no such correction at the next-to-leading order,
if we use Lorentz-invariants (in place of angles)
in the formulation\cite{tes2}\cite{smi}.

The value of the anomalous dimension $\gamma$
is around 0.20 for $\alpha_s\simeq 0.12$ (around the LEP energies).
The correction at $\theta =\pi /3$, for example, is about 5.7\%.
It is comparable to size of the uncertainties
from the yet uncalculated higher order corrections.

\newpage
\noindent {\large \bf FIGURE CAPTIONS}

\vspace{5 mm}

\noindent {\large Fig.1} \hspace{1 mm}
The lowest order diagrams for one-gluon emission cross section.

\vspace{5 mm}

\noindent {\large Fig.2} \hspace{1 mm}
The diagram for the soft-gluon emission (at a momentum $k_2$)
from a narrow jet of $q+g$.
We assume $k_2\cdot p_1\ll k_1\cdot p_1$.

\vspace{5 mm}

\noindent {\large Fig.3} \hspace{1 mm}
Four angular regions in the AO approximation.
The initial energetic $q\bar q$ is produced at the rectangle.
There is no soft-gluon emission in the region VI.

\end{document}